\documentclass[sigconf]{acmart}

\usepackage{booktabs} 
\usepackage{listings}
\usepackage{xspace, balance}

\renewcommand\footnotetextcopyrightpermission[1]{} 
\setcopyright{none}
\makeatletter
\def\@copyrightspace{\relax}
\makeatother
\setcopyright{none}

\newcommand{\projectname}{\textsc{TensorSCONE}\xspace}

\newcommand{\commentfontsize}{\fontsize{7}{8}\selectfont}

\newcommand{\myparagraph}[1]{\smallskip \noindent{\bf {#1}.}}

\newcommand{\out}[1] {}

\newcounter{codeLineCntr}

\reversemarginpar
\newif\ifnotes
\notestrue

\newcommand{\punt}[1]{}

\renewcommand{\eqref}[1]{Equation~(\ref{eq:#1})}

\newcommand{\proc}[1]{\ifmmode\mbox{\textsc{#1}}\else\textsc{#1}\fi}

  \newcommand{\func}[1]{\ifmmode\mathrm{#1}\else\textrm{#1}fi} %

\newcounter{remark}[section]

\begin{document}
\title{\textsc{TensorSCONE}:\\ A Secure TensorFlow Framework using Intel SGX}
\author{%
	{\em Roland Kunkel$^\dag$, Do Le Quoc$^\dag$$^*$, Franz Gregor$^\dag$$^*$, Sergei Arnautov$^*$, Pramod Bhatotia$^\ddag$, Christof Fetzer$^\dag$$^*$}\\
	{\small $^\dag$TU Dresden \quad $^*$Scontain UG \quad $^\ddag$The University of Edinburgh}\\
	{\small Technical Report, Feb 2019}
}

\date{}


\begin{abstract}
Machine learning has become a critical component of modern data-driven online services. Typically, the training phase of machine learning techniques requires to process large-scale datasets which may contain private and sensitive information of customers. This imposes significant security risks since modern online services rely on cloud computing  to store and process the sensitive data. In the untrusted computing infrastructure, security is becoming a paramount concern since the customers need to trust the third-party cloud provider. Unfortunately,  this trust has been violated multiple times in the past.

To overcome the potential security risks in the cloud, we answer the following research question: {\em how to enable secure executions of machine learning computations in the untrusted infrastructure?} 
To achieve this goal, we propose a hardware-assisted approach based on Trusted Execution Environments (TEEs), specifically Intel SGX, to enable secure execution of the machine learning computations over the private and sensitive datasets. More specifically, we propose a generic and secure machine learning framework based on Tensorflow, which enables secure execution of existing applications on the commodity untrusted infrastructure. In particular, we have built our system called \projectname from ground-up by integrating TensorFlow with  SCONE, a shielded execution framework based on Intel SGX. The main challenge of this work is to overcome the architectural limitations of Intel SGX in the context of building a secure TensorFlow system.  Our evaluation shows that we achieve reasonable performance overheads while providing strong security properties with low TCB.

\end{abstract}

\maketitle

\section{Introduction} 
\label{sec:introduction}
Machine learning has become an increasingly popular approach for solving various practical problems in data-driven online services~\cite{taigman2014deepface, bennett2007netflix,foster2014machine, deepmind_health}. While these learning techniques based on private data {\em arguably} provide useful online services, they also pose serious security threats for the users.  Especially, when these modern online services use the third-party untrusted cloud infrastructure for deploying these computations.



In the untrusted computing infrastructure, an attacker can compromise the confidentiality and integrity of the computation.  Therefore,  the risk of security violations in untrusted infrastructure has increased significantly in the third-party cloud computing infrastructure~\cite{Santos2009}.  In fact, many studies show that software bugs, configuration errors, and security vulnerabilities pose a serious threat to computations in the cloud systems~\cite{Gunawi_bugs-in-the-cloud, Baumann2014, Santos2012}.  Furthermore, since the data is stored outside the control of the data owner, the third-party cloud platform provides an additional attack vector.  The clients currently have limited support to verify whether the third-party operator, even with good intentions, can handle the data with the stated security guarantees~\cite{pesos, Vahldiek-Oberwagner2015}.

To overcome the security risks in the cloud, our work focuses on securing machine learning computations in the untrusted computing infrastructure.  In this context, the existing techniques to secure machine learning applications are limiting in performance~\cite{graepel2012ml}, trade accuracy for security~\cite{du2003using} or support only data classification \cite{bost2015machine}. 
Therefore, {\em we want to build a secure machine learning framework that supports existing applications while retaining accuracy, supporting both training and classification, and without compromising the performance}.

%
%

To achieve our design goals, we aim to leverage the advancements in trusted execution environments (TEEs), such as Intel SGX~\cite{intel-sgx} or ARM TrustZone~\cite{arm-trustzone}, to build a secure machine learning system.  In fact, given the importance of security threats in the cloud, there is a recent surge in leveraging TEEs for shielded execution of applications in the untrusted infrastructure~\cite{Baumann2014,arnautov2016scone, tsai2017graphene, shinde2017panoply, Orenbach2017}. {\em Shielded execution} aims to provide strong confidentiality and integrity properties for applications using a hardware-protected secure memory region or {\em enclave}.  

While these  shielded execution frameworks provide strong security guarantees against a powerful adversary, these systems have not been designed in the context of securing an existing machine learning framework, such as TensorFlow~\cite{abadi2016tensorflow}. To bridge this research gap, we propose \projectname
, a secure machine learning framework that supports both training and classification phases, while providing all three important design properties: {\em transparency},  {\em accuracy}, and {\em performance}. More specifically, we base our design on \textit{TensorFlow}, a widely-used machine learning framework. Our design builds on integrating TensorFlow with the SCONE~\cite{arnautov2016scone} shielded execution framework based on Intel SGX. 

However, it is not that straightforward to build a secure machine learning system
using shielded execution since it requires supporting unmodified applications without compromising the performance. Especially, we need to address three architectural limitations of shielded execution in our context: Firstly, the secure
enclave physical memory region is quite limited in size, and incurs high performance
overheads for memory accesses due to secure paging. This implies that we need to ensure that the memory footprint of TensorFlow library is minimal. Further, since the input dataset cannot fit into the enclave memory, we need to ensure that the data can be securely stored in the untrusted host memory and the untrusted file system.  Secondly, the syscall-based I/O operations are quite expensive
in the context of shielded execution since the thread executing the
system call has to exit the enclave, and perform a secure context switch, including TLB flushing, security checks, etc.  Therefore, it is  clearly not well-suited for building a secure intelligent application that requires frequent I/O calls. Lastly, since the TEE cannot give any security guarantees
beyond the enclave memory, we need to design mechanisms for extending the trust  to a distributed computing environment, which requires extending the trust over the network interface.

To overcome these design challenges, we present \projectname, a secure machine learning framework for the untrusted infrastructure. 
Overall, we make the following contributions.
\begin{itemize}

\item We have designed and implemented \projectname as the end-to-end system based on TensorFlow and SCONE that allows secure execution of the existing unmodified TensorFlow applications without compromising the accuracy. 
 
\item We optimized the performance to overcome the architectural limitation of Intel SGX in the context of machine learning workloads.

\item We evaluated \projectname with several microbenchmarks and a real world application. Our evaluation shows that \projectname achieves reasonable performance overheads, while providing strong security with low TCB.

\end{itemize}

 An early version of \projectname is already upstreamed and available as part of the SCONE framework  for production use: \href{https://sconedocs.github.io/tensorflowlite/}{https://sconedocs.github.io/tensorflowlite/}

\section{Background and Threat Model}
\label{sec:background}

\subsection{Intel SGX and Shielded Execution}
Intel Software Guard Extension (SGX) is a set of x86 ISA extensions for Trusted Execution Environment (TEE)~\cite{cryptoeprint:2016:086}. SGX provides an abstraction of secure \emph{enclave}---a hardware-protected memory region for which the CPU guarantees the confidentiality and integrity of the data and code residing in the enclave memory. The enclave memory is located in the Enclave Page Cache (EPC)---a dedicated memory region protected by  an on-chip Memory Encryption Engine (MEE). The MEE encrypts and decrypts cache lines with writes and reads in the EPC, respectively. Intel SGX supports a call-gate mechanism to control entry and exit into the TEE. 

{\em Shielded execution} based on Intel SGX aims to provide strong confidentiality and integrity guarantees for applications deployed on an untrusted computing infrastructure~\cite{Baumann2014,arnautov2016scone, tsai2017graphene, shinde2017panoply, Orenbach2017}. Our work builds on the SCONE~\cite{arnautov2016scone} shielded execution framework.
In the SCONE framework, the applications are statically compiled and linked against a modified standard C library (SCONE libc). In this model, application's address space is confined to the enclave memory, and interaction with the  untrusted memory is performed via the system call interface. In particular, SCONE runtime provides an {\em asynchronous system call} mechanism~\cite{flexsc}  in which threads outside the enclave asynchronously execute the system calls. 
Furthermore, it ensures memory safety~\cite{intel-mpx} for the applications running inside the SGX enclaves~\cite{kuvaiskii2017sgxbounds}.  Lastly, SCONE provides an integration to Docker for seamlessly deploying container images.

\subsection{Machine Learning using TensorFlow}

\begin{figure}
    \centering
        \includegraphics[width=0.48\textwidth]{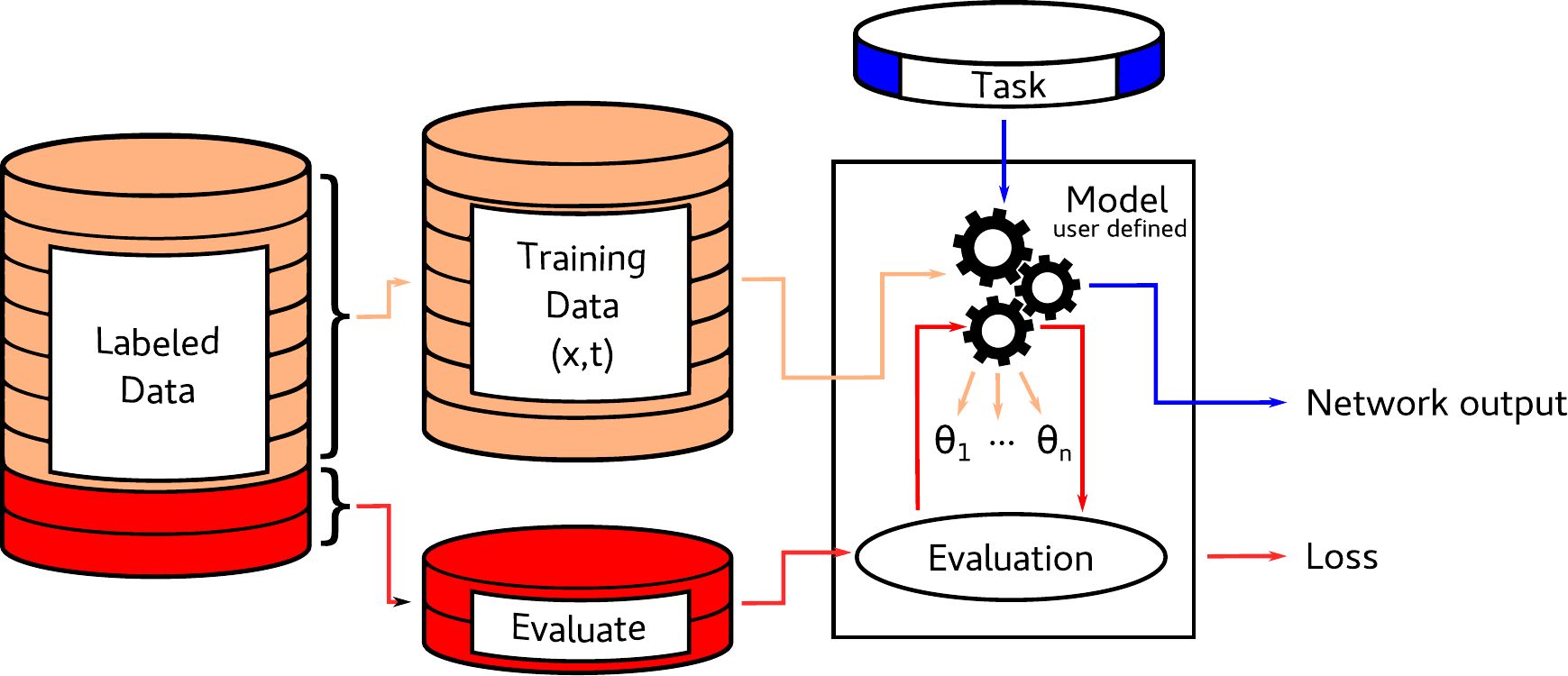}
    \caption[Machine learning approach]{Overview of the machine learning approach. The user provides a model according to the task and training data divided into training and evaluation data. The training data is used to update the parameters $\theta$ of the model in order to fit the training data, while the evaluation data is used to judge the solution. Data, which is to be classified later, is fed into the model the same way.}
    \label{fig:machine-learning-flow}
\end{figure}

Machine learning approaches aim to find solutions to problems by automatically deducing the required domain knowledge from example datasets~\cite{simeone2017brief}. Particularly, statistical models are leveraged to allow an information retrieval system to generalize and learn domain knowledge in order to solve a specific task. Broadly speaking, the machine learning approaches can be distinguished: supervised, unsupervised and reinforcement learning.
All forms have in common that they require data sets, a defined objective function, a model and a way to update the model according to new inputs. In our work, we focus on supervised learning, but our approach is generalizable to  the other two types. An overview of the process can be seen in Figure~\ref{fig:machine-learning-flow}.

To generalize the machine learning approach for masses, Google proposed TensorFlow~\cite{abadi2016tensorflow} as a machine learning framework designed for heterogeneous distributed systems. TensorFlow requires the user to first define a directed graph consisting of nodes representing operations on incoming data. Nodes perform computations on different levels of abstraction such as matrix multiplication, pooling or reading data from disk.
Nodes can also have an internal state, depending on their type. The stateful operations called {\em Variable} which contains mutable buffer used to store shared and persistent state across multiple iterations. The data flows along the directed edges  in the graph called {\em Tensors} --- the n-dimensional abstraction of matrices.

After defining the graph, the user can perform calculations in the graph by starting a session and running the previously defined operations. TensorFlow uses a dataflow model for calculations, in which an output of one operation(i.e., a node) becomes the input for another operation.



Currently, TensorFlow supports distributed training, allowing part of the graph to be computed on different physical devices. TensorFlow can be deployed on mobile devices, single personal computers, as well as computer clusters, by mapping the computation graph on available hardware.  This framework allows multiple devices to be used to train a model, with parameters and weights being shared between them. Each iteration of the execution over the computation graph is called a {\em step}.

TensorFlow Lite~\cite{tensorflow-lite} is a feature-reduced version of TensorFlow, designed for mobile and embedded devices.
Optimization for mobile devices is achieved by running a mobile-optimized interpreter that keeps the load at a lower level and by keeping the overall binary size smaller when compared to full TensorFlow, among other measures.

The number of available operations for defining a graph is reduced to achieve a smaller memory footprint of the resulting binary.
Currently, TensorFlow Lite does not support training. To use the framework, a model must first be training with the full version of TensorFlow and then exported and converted to a special TensorFlow Lite model format. This format can then be used from the TensorFlow Lite API for inference.


\subsection{Threat Model}
We  aim to protect against a very powerful adversary even in the presence of complex software layers in the virtualized cloud computing infrastructure~\cite{Baumann2014}. In this setting, the adversary can control the entire system software stack, including the OS or the hypervisor, and is able to launch physical attacks, such as performing memory probes. Even under the extreme threat model,  our goal is to guarantee data integrity, confidentiality, and freshness. 
Data freshness property ensures that the data is recent and there is no old state of data has been replayed. 
We also provide bindings with Pesos~\cite{pesos}, a secure storage system to protect against rollback attacks~\cite{Parno2011} on the data stored beyond the secure enclave memory.
Further, since we provide memory safety using SGXBounds~\cite{kuvaiskii2017sgxbounds}, \projectname is resilient to an important class of code-reuse attacks on SGX~\cite{code-reuse}. 

However, we do not protect against side-channel attacks based on cache timing and speculative execution~\cite{foreshadow}, and memory access patterns~\cite{xu2015controlled, hahnel2017high}. Mitigating side-channel attacks is an active area of research~\cite{varys}. 
Lastly, we do not consider denial of service attacks since these attacks are trivial for a third-party operator controlling the underlying infrastructure~\cite{Baumann2014}. 
Lastly, we assume that the adversary cannot physically open the processor packaging to extract secrets or corrupt the CPU system state.

\section{Design} 
\label{sec:design}
In this section, we present the design of \projectname. 
%
\subsection{Overview} \label{sec:general-architecture}
%

\begin{figure}
    \centering
                \includegraphics[width=0.33\textwidth]{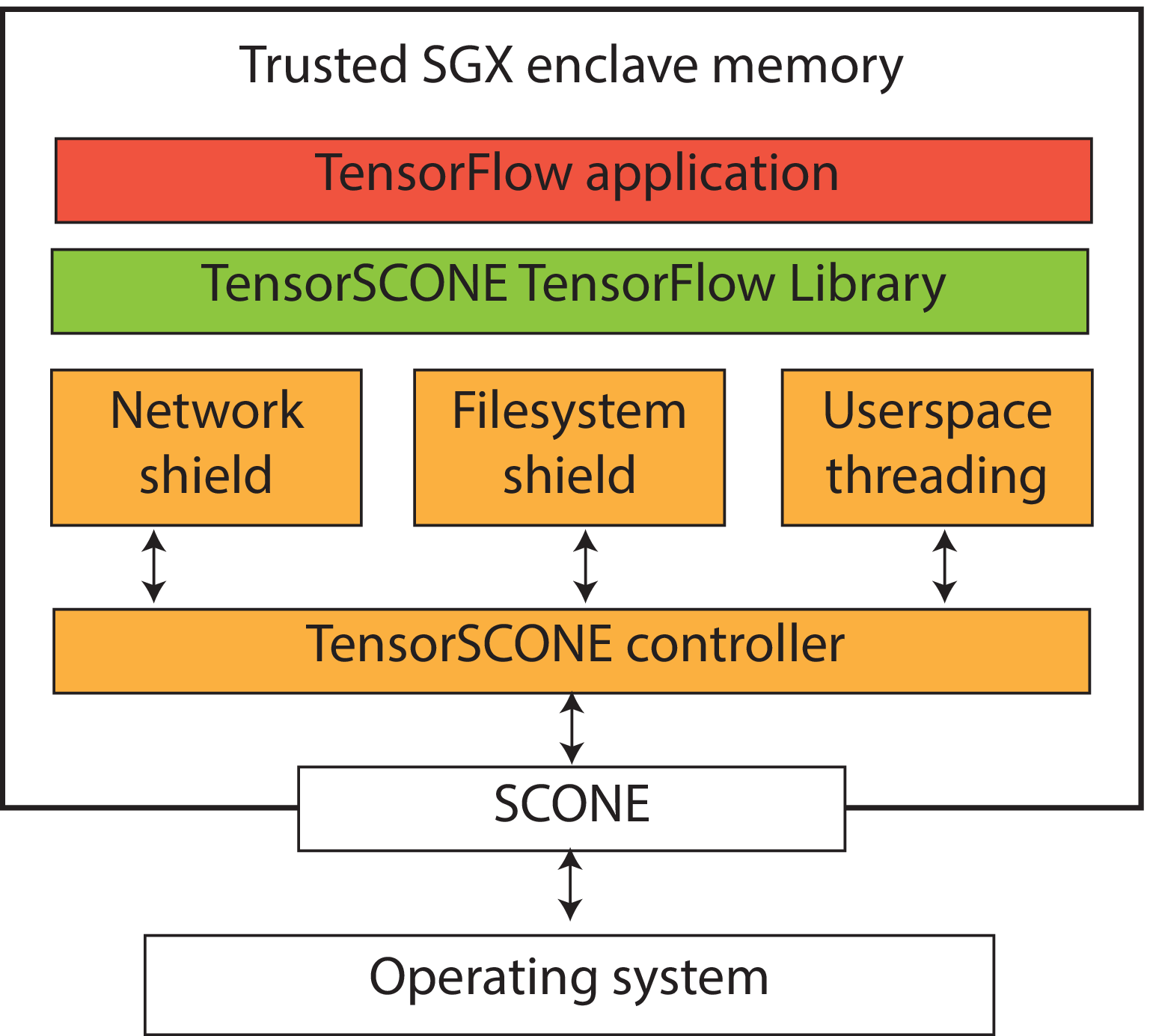}
    \caption[\projectname architecture]{
        The architecture of \textit{\projectname}.}
    \label{fig:tensorscone-arch}
\end{figure}

%
%
At a high-level, our strawman design consists of the Tensor machine learning framework, which is secured by the hardware-assisted trusted execution environment (TEE). We base our design on TensorFlow and TensorFlow Lite for supporting the machine learning workloads. TensorFlow Lite has the additional advantage of having a smaller memory footprint. The TEE we choose for our work is Intel \textit{SGX}. Using Intel SGX directly to secure an application requires rewriting the application specifically for SGX, which can be complex. We therefore use \textit{SCONE} as an additional layer that allows access to SGX features with fewer changes to application code.
While there are other options available, we choose SCONE, because of the relatively small extra work required to run an application and comparatively small overhead compared to other available options. In particular, we integrated TensorFlow with the SCONE shielded execution framework. Figure ~\ref{fig:tensorscone-arch} presents the general architecture of \projectname.

As the first step, when a user deploys an application on a remote host, the user can only be certain the correct application is running untampered with, if the application is running inside a TEE, and the identity of the application has been proven.
The attestation step is therefore an integral part of the life cycle of common TEEs.
SGX offers attestation mechanisms as well. However, SCONE, as an additional layer between SGX and application, exposes a standardized interface for performing remote attestation that is independent of a particular application.  We therefore leverage the SCONE framework to provide the remote attestation mechanism to verify the proof of integrity and security of the application running on a distributed cluster of machines inside the cloud. 

The communication channel for data exchange between \projectname and the user must be private and secure.
The security of this channel must be end-to-end protected, starting inside the TEE and terminating at the user side.
The \textit{TLS} protocol offers all properties we need.
Finally, we use Docker for easier distribution of our system.  In particular, we use the SCONE infrastructure for securely distributing configuration for the containerized applications.


\myparagraph{Design goals}  Our primary design goal is to achieve strong confidentiality and integrity properties for the secure execution of machine learning applications. By confidentiality, we mean that all data  handled by the machine learning framework and the machine learning framework code itself may not be disclosed to or obtainable by an unauthorized party. By integrity, we mean that modifications of the data handled by \projectname that were done by an unauthorized party must be detectable and should not compromise the internal state and functioning. In addition, while designing a practical system based on the strawman design, we aim to achieve the following design goals.

%
%

\begin{itemize}
\item{\em Transparency:}
The secure  framework must offer the same interface as the unprotected framework, and should unmodified existing applications based on TensorFlow. 

\item{\em Performance:}
We aim to impose as little overhead as possible when adding security to the machine learning framework.

\item{\em Accuracy:} We do not aim to trade-off accuracy for security. Accuracy will be the same of the native TensorFlow framework as when using no security protection.

\end{itemize}

%
%
\subsection{Detailed Design} \label{sec:conrete_arch}

The design of \projectname is composed of two components: (a) the \projectname controller that provides the necessary runtime environment for securing the TensorFlow library, and (b) \projectname TensorFlow library that allows deploying unmodified existing TensorFlow applications. We next describe these two components in detail.



%

%

\subsection{\projectname Controller}

\myparagraph{General architecture} The \projectname controller is based on the SCONE shielded execution framework. 
The \projectname controller runs inside a Docker container \cite{merkel2014docker}.
No changes to the Docker engine is required.
Inside the enclave, the controller provides a runtime environment for TensorFlow, which includes the \textit{network shield}, the \textit{file system shield}, user-level threading. 
These subsystems are required in order to transparently support unmodified existing TensorFlow applications inside the SGX environment.
Data that is handled through file descriptors is transparently encrypted and authenticated through the shields.
The shields apply at each location where an application would usually trust the operating system, such as when using sockets or writing files to disk.
The shields perform sanity checks on data passed from operating system to enclave to prevent Iago attacks~\cite{Checkoway2013}.
More specifically, these checks include bound checks and checking for manipulated pointers.
This protection is required to fulfill the goal of not requiring the application to deal with untrusted systems.

\myparagraph{File system shield} The file system shield protects confidentiality and integrity of data files.
Whenever the application would write a file, the shield either encrypts and authenticates, simply authenticates or passes the file as is.
The choice depends on user-defined path prefixes, which are part of the configuration of an enclave.
The shield splits files into chunks that are then handled separately.
Metadata for these chunks is kept inside the enclave, meaning it is protected from manipulation.
The secrets used for these operations are different from the secrets used by the SGX implementation.
They are instead configuration parameters at the startup time of the enclave.

%
%

\myparagraph{Network shield}  TensorFlow applications do not inherently include end-to-end encryption for network traffic.
Users who want to add security must apply other means to secure the traffic, such as a proxy for the Transport Layer Security (TLS) protocol.
According to the threat model however, data may not leave the enclave unprotected, because the system software is not trusted.
Network communication must therefore always be end-to-end protected.
Our network shield wraps sockets, and all data passed to a socket will be processed by the network shield instead of the system software.
The shield then transparently wraps the communication channel in a TLS connection on behalf of the user application.
The keys for TLS are saved in files and protected by the file system shield.
%

\myparagraph{User-level threading}
Enclave transitions are costly and should therefore be avoided when possible.
Many system calls require a thread to exit userspace and enter kernel space for processing.
To avoid thread transitions out of enclaves as much as possible, the controller implements user space threading.

When the OS assigns a thread to an enclave, it first executes an internal scheduler to decide, which application thread to execute.
These application threads are then mapped to SGX thread control structures.
When an application thread blocks, the controller is run again to assign the OS thread to a different application thread instead of passing control back to the operating system.
In this way, the number of costly thread transitions is reduced.
When no application thread is ready for execution, the OS either backs off and waits inside the enclave, or outside, depending on the time required for an enclave transition.
A side effect of this user-level threading scheme is that the controller does not require more OS threads than CPUs available to achieve full CPU utilization, which is usually the case for applications running under a conventional OS.

\subsection{\projectname TensorFlow Library}

Machine learning applications consist of two major steps.
In the first step, the model is trained, and thereafter, the model is employed for classification or decision tasks. We next explain the two stages of the workflow: training process and classification process.

\myparagraph{Training process} For the training process, we use the full version of TensorFlow.
Training in TensorFlow is usually performed on acceleration hardware such as {GPU}s and distributed across multiple machines.
However, the \projectname controller requires SGX which is only available for CPUs. 
We therefore only support training on \textit{CPU}.
This limitation reduces the performance of the training process, but additional security is added.

The \projectname controller  allows easy distribution of the application in form of docker images.
The training instances of \projectname can be distributed on multiple nodes, each running separate SGX hardware.
The network shield applies transparent protection of the communication channel between instances.
Scaling on the same instance, that is, on the same CPU is possible, but does decrease relative performance, because the limiting factor in our environment is EPC size, which is fixed for each CPU.
Only horizontal scaling with more instances can increase performance.

The system calls required by TensorFlow can be seen in Table \ref{tab:tensorflow-syscalls}.
Again, most time is spend handling \textit{futex}.
\begin{table}[t]
\centering
\commentfontsize
\begin{tabular}{|l||l|l|l|l|l|}
\hline
syscall & futex   & sched\_yield & nanosleep & munmap  & brk      \\ \hline
\hline
time (s) & 6421    & 448          & 441       & 0.09    & 0.05 \\ \hline
time (\%) & 87.83   & 6.14          &  6.04   & \textasciitilde0 & \textasciitilde0 \\ \hline
\end{tabular}
\caption{System calls required by TensorFlow for training 1,000 steps of the Cifar-10 model, which took about 19:30 min of real time (including overhead from the measurement).}
\label{tab:tensorflow-syscalls}
\end{table}

\myparagraph{Classification process} 
The main reason for dividing the classification and training process in our design is that we can use different TensorFlow variants for each step. 
SCONE imposes less overhead, if applications have a smaller memory footprint, because the limited \textit{EPC} size is the major bottleneck.
TensorFlow Lite has a smaller memory footprint because it targets mobile devices.
The drawback is however that it cannot perform training by design.
Therefore, we can only use it for classification.

When protecting TensorFlow Lite with SCONE, the framework uses the SCONE C library instead of the common system library.
The internals of TensorFlow Lite do not otherwise require change, as long as the interface of the SCONE C library is fully compatible.
The most common system calls required by TensorFlow Lite can be seen in Table \ref{tab:lite-syscalls}.
Most time is spent to handle \textit{futex}, which is a userspace lock that does not require switching to kernel space.
\begin{table}[t]
\centering
\commentfontsize
\begin{tabular}{|l||l|l|l|l|}
\hline
syscall & futex   & read & munmap & write\\ \hline
\hline
time (s) & 4.68    & 0.06 & 0.004 & \textasciitilde0\\ \hline
time (\%) & 98.73   & 1.18 &  0.09 & \textasciitilde0\\ \hline
\end{tabular}
\caption{System calls required by TensorFlow Lite for classifying 1,000 images, which took about 7:30 min of real time (including overhead from the measurement).}
\label{tab:lite-syscalls}
\end{table}
The interface for using the classification method of \projectname is the same as for TensorFlow Lite.
Graph definitions created for TensorFlow Lite are compatible.

\section{Implementation} 
\label{sec:implementation}
We next explain how we put the design of \projectname into practice. \projectname is upstreamed and integrated with the SCONE framework.
%

\subsection{Training Process} \label{sec:tensorflow-full-with-scone}
The typical user of TensorFlow uses the Python API for defining and training graphs, because it is the richest API.
Using Python with SCONE would impose additional complexity because it requires the \textbf{d}ynamic \textbf{l}ibrary \textbf{open} (\texttt{dlopen}) system call for imports.
As the name implies, \texttt{dlopen} dynamically loads libraries during runtime of a program.
However, \textit{SGX} does not allow an enclave to be entered by a thread, unless it has been finalized according to the procedures of enclave creation. 
A library that is dynamically loaded would therefore not be represented in the enclave's attestation hash. 
Consequently, \texttt{dlopen} is disabled by default for SCONE applications.
The designer of a SCONE service can decide to allow \texttt{dlopen} by configuring the SCONE environment accordingly.
Doing so requires further deliberation, if the security is not to be compromised.
The standard way for handling this case is to authenticate loaded libraries during runtime through the file system shield.

The TensorFlow repository offered convenience scripts for creating Python packages for distributing TensorFlow.
Compiling the packages with the SCONE failed, because parts of the Python package require the \texttt{fork} system call, which was not yet fully available in SCONE during the time of this work.
The implementation for \texttt{fork} in SCONE was available, but not yet part of the SCONE release.

We therefore decided to support only the C++ API for TensorFlow.
The C++ version covers the low-level API of TensorFlow, meaning many convenience features such as estimators or monitored training are not available.
However, the TensorFlow core is written in C++, and the C++ API is feature complete in a sense that everything that can be done with the Python API can also be done with the C++ API, but requires more verbose source code.

There is one approach that let us use the convenience of the Python API for the definition of the graph.
TensorFlow allows exporting graphs and parameters, such as learned biases that were created in the current session.
Graph definitions and checkpoints containing the parameters can later be imported by another program.
Importing and exporting are available in both the C++ and the Python API, and they use interchangeable exchange formats.
The user can therefore define a graph with the more high level Python API, including data inputs, and later import and run it with C++.
If the application does not by default already export its model with a named interface, changes are required to the original program, so that either the name of operations in the graph can be known, or an interface is defined.
%

\begin{figure}
    \centering
        \includegraphics[width=0.48\textwidth]{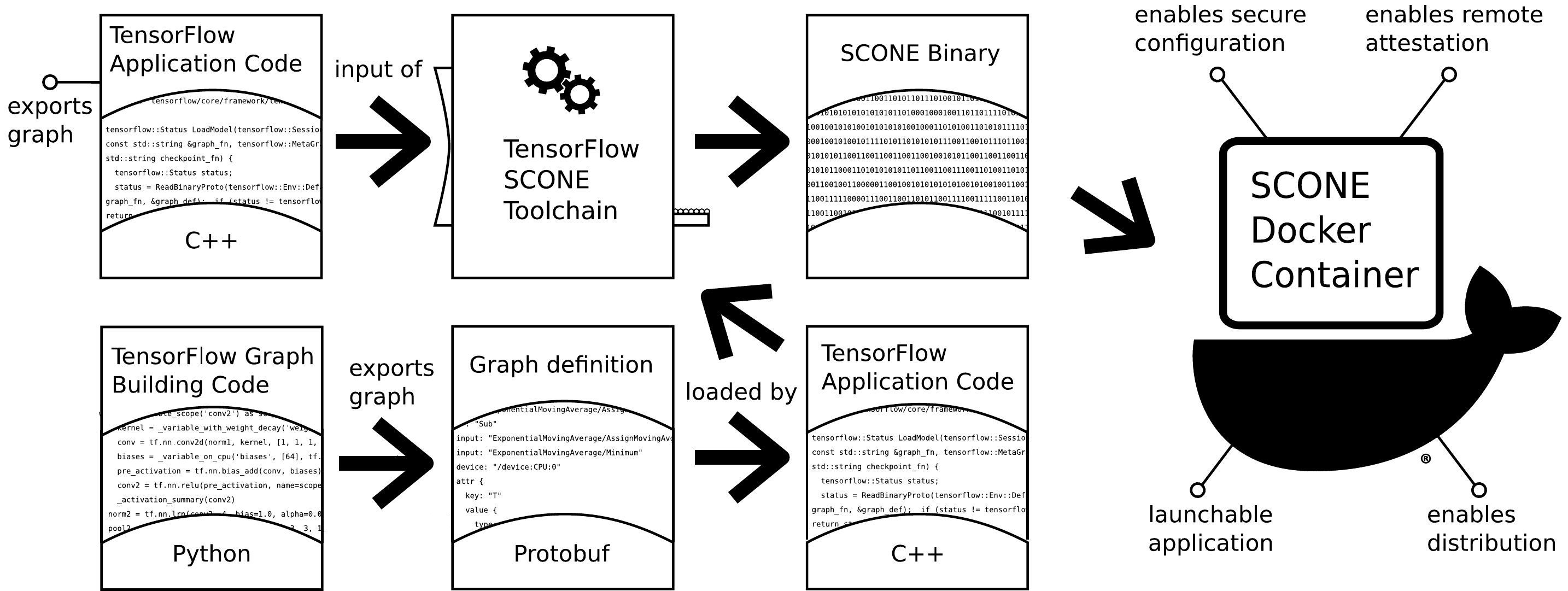}
    \caption[\projectname interface]{
        Interface and flow of \textit{\projectname}.
        The user either provides a C++ program building and running a TensorFlow graph, converts it with a \projectname toolchain to a binary that can be run in a SCONE container.
        Alternatively, a Python program exporting a graph, and a corresponding C++ program for training the graph can be provided.}
    \label{fig:tensorscone-interface}
\end{figure}
For the training process, we used the full version of TensorFlow, not to be confused with TensorFlow Lite.
A graph definition must be provided by the user in form of a graph \textit{frozen} by a script packaged together with TensorFlow, when using either the Python or C++ API.
If the user has used the C++ API for the definition, the full source definition of the graph can also be used.

A frozen graph can be created from a graph definition exported from the Python script that defines the graph in the \textit{Protocol Buffers} (\cite{protobuf}) exchange format.
A checkpoint file containing all values of a graph that are not part of the graph definition, such as weights, biases and counters can be exported as well.

Alternatively, the graph can also be exported as a blank slate without any initialized internal values.
The initialization can then be done inside the \projectname environment, which is useful if a user wants to train the graph protected by SGX for the entire training process.
The initialization operations are required when using the Python API and are therefore usually part of the exported graph.

The user must also provide the inputs for training, such as a set of annotated images.
The code written for classification must use the TensorFlow Lite API.
If the confidentiality of the training material is required, the file system shield of SCONE can be used.
The configuration is straightforward and can be easily distributed through SCONE mechanisms.
An overview of the interface and usage workflow can be seen in Figure \ref{fig:tensorscone-interface}.

\subsection{Classification Process} 
\label{sec:tensorflow-lite-implementation}
We implemented our design for the classification process by running the TensorFlow Lite framework with SCONE.
For testing, we used the C++ \textit{API}.
We first ensured that TensorFlow Lite compiles with the musl C library on Alpine Linux \cite{alpine_linux}, because SCONE uses a modified version of the musl library.
The Docker containers built for distributing SCONE are also based on Alpine.
Alpine Linux is a lightweight Linux distribution that uses the musl C library by default.

Musl is designed to be compatible with \textit{glibc} without changes to the application.
In practice however, changes can be necessary as we will show.
\textit{Identical code folding}(ICF) is a compiler or linker feature that eliminates identical function bodies at compile or link time in order to reduce the binary size.
It is currently supported by \textit{gcc} and the gold linker, but not by the musl linker or the compiler wrapper for musl.
We therefore removed the \textit{ICF} option for the binary targets in the TensorFlow source tree.
Compiling the TensorFlow framework with and without ICF provides similar binary sizes.
Therefore, the performance cost when deactivating ICF will also be minimal.

TensorFlow also uses \texttt{backtrace} by default.
This library is specific for glibc.
We therefore could not use it directly with musl.
There are unstable alternatives and stubs available to replace backtrace for Alpine Linux programs, but we decided to disable the option entirely to avoid pulling more unstable dependencies into the project \cite{alpine_backtrace}.
One way for disabling dependencies is to add compile guards to conditionally only compile a dependency into the binary, when certain circumstances are met, such as compiling against the musl libc.
I

The TensorFlow source uses \textit{Bazel} as a build tool \cite{bazel}.
Bazel was first, like TensorFlow, developed internally by Google and released into public in 2015.

Integrating SCONE into an application in the simplest case merely requires compiling the application with a SCONE-specific wrapper for gcc, but it can be more complex in practice.
Bazel allows the configuration of custom build chains that can be specified by a command line switch when invoking Bazel for builds.
The configuration of the toolchain we created is described in section \ref{sec:scone-toolchain}.

SCONE uses environmental variables for specifying the amount of memory available for heap and stack when running the enclave.
The default stack size was enough when running TensorFlow Lite.
For the heap size, a minimum of about 220 \textit{MB} had to be set for the classification of up to a tested amount of 1,000 images.
After classification, images were not kept in memory, meaning higher classification counts do not need more memory.

We confirmed the correctness of the implementation by comparing the classification results delivered by \projectname with the results of native TensorFlow Lite.
We checked the values of the top four labels.
They had the exact same percentage and order in both cases, when classifying images randomly picked from the web.
We could therefore be sure that classification with \projectname works correctly.

%
To the best of our knowledge, there is no standalone version of TensorFlow Lite available, meaning a user of TensorFlow Lite needs to build their application inside the TensorFlow source folder, with dependency targets set to TensorFlow Lite.
This is a major limitation for existing projects.
Bazel also does not link library targets unless a binary target is created, which means TensorFlow Lite cannot be easily released from the source tree by compiling all libraries, and move them to the system's include directories.

We added compile targets that force linking as a workaround.
The libraries could then be moved to other projects along with the header files, and used as third party dependencies.
With this, we wrote a classifier service from scratch.
The service takes classification requests via network, and uses TensorFlow Lite for classification.
The classifier service serves as a proof of concept.
It is not used for the performance tests, because the service also implements a custom communication protocol, which is out of scope for the measurements.

For testing, we used an example available in the TensorFlow Lite source, which takes its inputs from the hard drive and prints the classification results to console.
Presumably, for benchmarking purposes, the authors included a command line option to run the classification subroutine for a certain number of times.
We used this option to simulate batch processing of many images at once, without requiring to restart enclaves every run.

\subsection{\projectname Toolchain} \label{sec:scone-toolchain}
Our solution to compile TensorFlow applications for SGX was to setup a new toolchain for SCONE inside the TensorFlow repository.
Tools for cross compilation for ARM processors or CUDA capable GPUs were already available in the TensorFlow repository.
A toolchain definition requires the configuration of paths to specific tools such as the linker \texttt{ld}, GNU Compiler Collection \texttt{gcc} and other tools.
Furthermore, the correct include paths for system libraries and other options must be configured.
For SCONE, this included dynamic and position independent linking.
Defining a toolchain was the cleanest solution, because it allows users to easily switch between compilers without side effects.
It was also necessary to define a toolchain, when different tools are required for different parts of the project.
The Protobuf compiler for example could be used with the native toolchain.  
Another solution for compiling TensorFlow Lite with SCONE was to set the environment variables \texttt{CC} and \texttt{CXX} to the SCONE C and C++ compilers. This solution only worked when the Bazel output path was set to a custom path outside of the build directory.
According to the Bazel documentation, this is intended for debugging purposes. When using this approach, the \texttt{LD\_LIBRARY\_PATH} environment variable needed to be set to the output folder. This solution should be avoided, because tools required only by the compilation host will also be compiled with the SCONE toolchain, making cross compilation impossible. 


\section{Evaluation}
\label{sec:evaluation}
In this section, we first present the experimental setup. Thereafter,  we evaluate a real world application of \projectname by training the Cifar-10 model. 

\subsection{Experimental Setup} 
\label{sec:setup}
For all experiments, we used servers running Ubuntu Linux with a 4.4.0 Linux kernel, equipped with an Intel\textcopyright{} Xeon\textcopyright{} \textit{CPU} E3-1280 v6 at 3.90\textit{GHz} and 64 \textit{GB} main memory.
This processor was released in early 2017 and supports SGX.
The \textit{gcc} release for compiling the \textit{glibc} versions was 5.4.0.
The gcc release on which SCONE builds is 7.3.0.
Before the actual measurements, we warmed up the machine by running at full load with IO heavy operations that require swapping of \textit{EPC} pages.
We performed measurements for classification and training both with and without the file system shield.
For full end-to-end protection, the file system shield was required.

\subsection{TensorFlow Application} 
\label{sec:tensorflow-full-usage}
For demonstrating the usage of \projectname, we ported a training application written in Python to the TensorFlow C++ \textit{API} and ran it with \projectname.
The training application trains the \textit{Cifar-10} data set and can be found in the \texttt{models}-project associated with TensorFlow \cite{tensorflow_models}.

\begin{figure}
    \centering
        \includegraphics[width=0.48\textwidth]{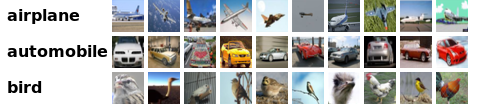}
    \caption[Cifar-10 data set]{The Cifar-10 image set contains 60,000 32x32 pixel images divided into 10 classes.} 
    \label{fig:cifar10}
\end{figure}

\myparagraph{Dataset: Cifar-10}
The Cifar-10 image set \cite{krizhevsky2009learning} is a labeled subset of a much larger set of small pictures of size 32x32 pixels collected from the Internet.
It contains a total of 60,000 pictures.
Each picture belongs to one of ten classes, which are evenly distributed, making a total of 6,000 images per class.
All labels were manually set by human labelers.
An example of classes and images can be seen in Figure \ref{fig:cifar10}.
Cifar-10 has the distinct advantage that a reasonable good model can be trained in a relatively short time.
The set is freely available for research purposes and has been extensively used for benchmarking machine learning techniques~\cite{xu2015empirical,hinton2012improving,he2016deep}.

\myparagraph{Model a.k.a. the graph}
The model we trained to classify the Cifar-10 data set is a convolutional neural network.
It consists of two convolutional layers, each followed by max pooling, and three fully connected layers with rectified linear unit (\textit{ReLU}) activation functions.
Softmax is used for deciding the classes.

For creating the graph, we used the Python API and exported the graph to a \textit{Protobuf} file.
We realized the training loop with the C++ API.
For this, we first imported the graph from the Protobuf file, and looped over the training function we defined in the Python part.
Queue runners, hooks and other parts that are automatically handled when training with the high level Python API had to be manually implemented.

The images are read from disk through a \texttt{FixedLengthRecordReader}.
This class enables the input of images into tensors.
It dedicates a full thread to this task.
Threads are automatically handled in Python with the \texttt{MonitoredTrainingSession} API.
For C++, we had to manually create a thread and bind the correct node to it.  

To enrich the inputs and mitigate overfitting of the model, the original images from the Cifar-10 data set are distorted.
This virtually enlarges the data set and allows for more generalized models.
First, only random 24x24 pixel crops are taken from each 32x32 original image.
Next, each image is randomly flipped, and has its brightness and saturation adjusted.
These image distortion functions are part of the \texttt{image}-API of the Python API.
The augmented data is randomly shuffled and fed into the model.
For training, the cross entropy loss of the softmax function is minimized.

\myparagraph{Application implementation} Defining a graph in TensorFlow creates nodes or \textit{operations}, at which the inputs are reshaped and produced into the outputs.
The flow of the data is realized through tensors.
Operations can be executed by passing the operation to a \texttt{run} call, both in the C++ and Python API of TensorFlow.
Training can be realized by defining an operation that performs a single gradient descent in the model, and then looping over that operation.
Both the C++ and Python API offer functions to import and export graphs and their parameters, allowing saving the trained graph, and using it for classification from a different application.

The most distinctive difference between both APIs for training is the \texttt{MonitoredTrainingSession}, which is offered for the Python API.
It allows convenient registration of hooks when running the session, initializes all variables and starts all worker threads involved in the graph, such as threads reading the images from disk or processing the shuffle queue.
This hides complexity from the user, because there is no need to keep track of required threads or names of operations.
These operations all have to be done manually with the C++ API, increasing both the amount of code that needs to be written.
It also makes it necessary to expose internals of the graph in form of the names of internal operations for filling queues and reading images.
The names can be set manually, if we have the possibility to author the graph definition.
When the source of the graph is not available, the names of the operations can be found using \textit{TensorBoard}, which is a tool for the visualization of TensorFlow executions \cite{tensorboard}.

The queue operations are blocking and had to be moved to different threads.
The C++ standard library offers \texttt{std::thread} objects, which can be used to run the operations concurrently.
They are also compatible with SCONE, which does by default offer four thread control structures  for execution, which is sufficient in our scenario. 

The initialization of variables can be performed in the Python code that builds the graph before exporting.
The initialized variables are then saved in a checkpoint file that is also exported. 
The values for the variables can then be loaded on the C++ side.
Alternatively, another operation that initializes the variables could be added to the graph and then run with the C++ API.

\subsection{Performance of the Classification Phase} 
\label{sec:tflite_microbenchmark}
We first evaluate the performance of \projectname for the classification process.

\subsubsection{Data set and model} 
\label{sec:tflite-data-set}
The data set we used for benchmarking \projectname consists of a single bitmap image.
For the latency measurement, we calculated the average over 1,000 classifications performed by TensorFlow Lite.
We conducted measurements for the native versions using glibc and musl, SCONE in simulation mode and SCONE running on SGX hardware.
\textit{Native} means execution that was performed without SGX and therefore also without SCONE.
Native executions ran, like the versions using SCONE, inside a Docker container.
The performance influence of Docker is therefore out of the equation.

The model we used for the classification of the images is \textit{Inception-v4} \cite{szegedy2017inception}.
It achieves 3.08\% top-5 error on the ImageNet classification challenge data set, making it state-of-the-art.
A pre-trained model is hosted in the TensorFlow repository \cite{tflite_models}.
A version already converted to the TensorFlow Lite Protobuf serialization format is also available.
We estimate that training \textit{Inception} from scratch would take months with the hardware we had at our disposal.

We manually checked the correctness of a single classification by classifying the image with the TensorFlow label image application involving no self-written code and running directly on the host without containerization.
We later compared the results to the ones provided by \projectname and could confirm that indeed the same result was produced.
\subsubsection{Results} \label{sec:tflite-results}
\begin{figure}
    \centering
        \includegraphics[width=0.4\textwidth]{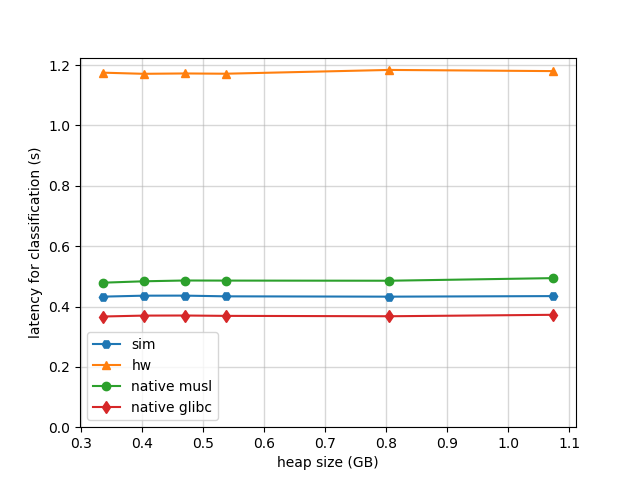}
    \caption[\projectname classification latency]{Latency in seconds when classifying images with TensorFlow Lite, using native execution with glibc, native execution with musl, SCONE with SGX hardware mode and SCONE with simulation mode.}
    \label{fig:tensorflow-lite-results}
\end{figure}

We present the results in Figure \ref{fig:tensorflow-lite-results}.
In this section, we explain the results and evaluate the influence of heap size and file system shield on the performance.

\myparagraph{\#1: Latency}
When compiled with glibc, TensorFlow Lite had the smallest latency (shown in red).
The version using musl (green line) had about 30\% greater latency.
Both C libraries excel in different areas, but glibc has the edge over musl in most areas, according to microbenchmarks \cite{clib_compare}, because glibc is tailored for performance, whereas musl is geared towards small size.
Because of this difference in goals, an application may be faster with musl or glibc, depending on the performance bottlenecks that limit the application.
Differences in performance of both C libraries must therefore be expected.
The deviation from the average of each single measurement was very low.

\myparagraph{\#2: Throughput}
The performance when executing with SCONE in simulation mode (blue) was slightly higher than compared with native musl execution.
One reason for this might be that SCONE handles certain system calls inside the enclave and does not need to exit to the kernel.
In simulation mode, the execution is not performed inside the enclave, but SCONE still handles some system calls in userspace, which can positively affect performance.
An analysis with the \texttt{strace} tool yields that some of the most costly system calls of \projectname are indeed system calls that are handled internally by the SCONE runtime. 

\begin{figure}
    \centering
        \includegraphics[width=0.4\textwidth]{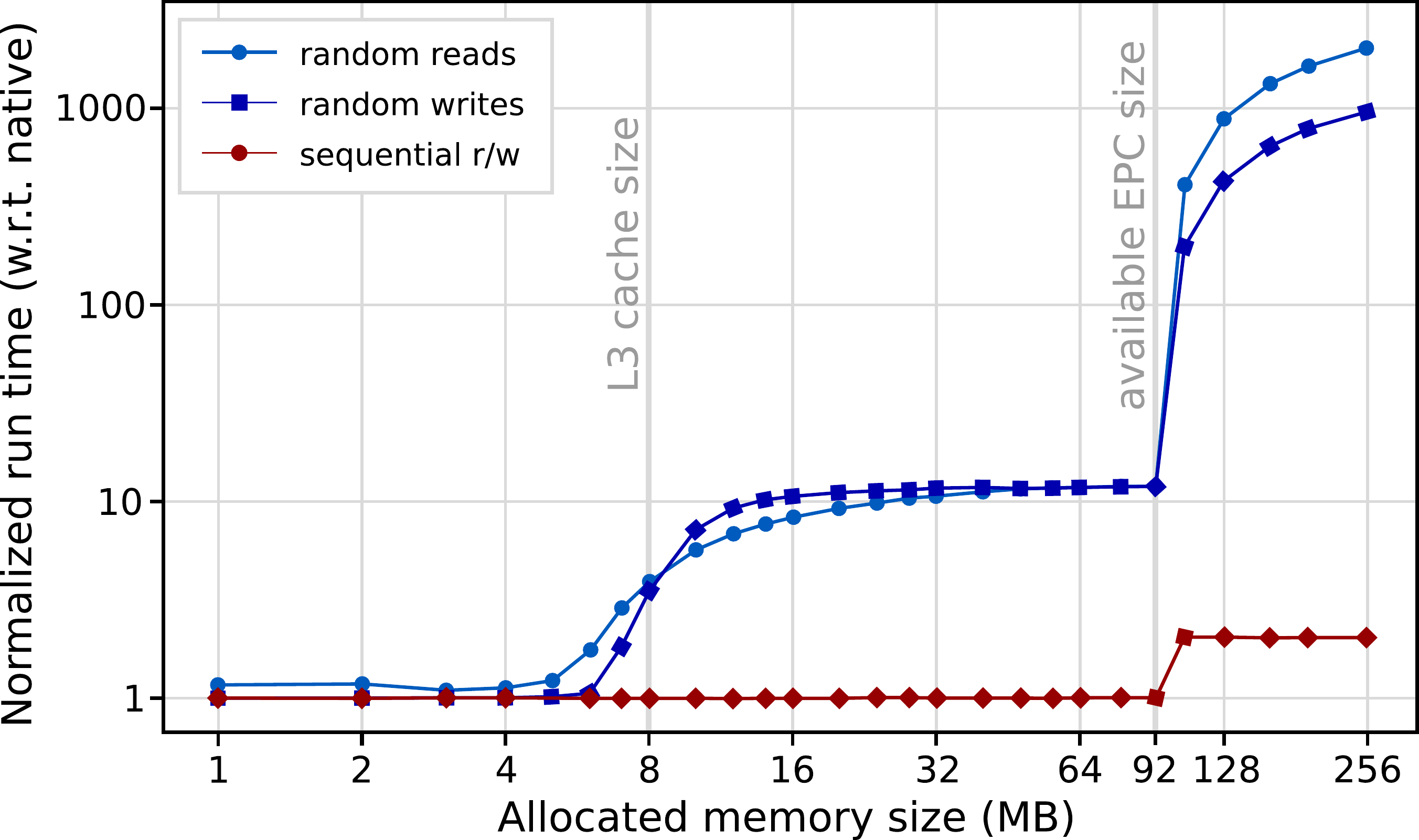}
    \caption[SCONE read/write microbenchmark]{The plot shows the required time for read and write calls in correlation to the total memory an enclave has allocated. A drastic decrease of performance of read and write calls occurs, when the touched memory resides outside the EPC. This serves as an explanation for the seen latency when classifying with \projectname (taken from \cite{arnautov2016scone}).}
    \label{fig:scone-microbenchmark}
\end{figure}

The time it takes to classify a single image also determines the throughput of \projectname.
We compare the throughput of different configurations in Table \ref{tab:classification-throughput}.
\begin{table}[]
\commentfontsize
    \centering
    \begin{tabular}{|l||l|l|l|l|}
    \hline
                     & native glibc & native musl & simulation & SGX   \\ \hline \hline
    throughput (1/s) & 2.685        & 2.081       & 2.314      & 0.848 \\ \hline
    compared to native & 1        & 0.78       & 0.86      & 0.32 \\ \hline
    \end{tabular}
    \caption[\projectname classification throughput]{Throughput of various configurations in classifications per second. We compare a version running native glibc, native musl, \projectname with simulated SGX, and \projectname with hardware SGX.}
    \label{tab:classification-throughput}
    \end{table}
The throughput of SCONE in hardware mode was about 0.32 of the throughput of native musl, as seen in Table \ref{tab:classification-throughput}.
While we expected some decrease in throughput, this result is considerably lower than 0.6 times native throughput, which is the lower bound of throughput compared to native execution that SCONE achieves, given by the authors of SCONE. 
A reason for getting subpar results might be the larger main memory area TensorFlow requires compared to the applications tested by the SCONE authors. Let's consider a microbenchmark presented in~\cite{arnautov2016scone}.
The key-value store Memcached for example, which achieved 1.2x the throughput of native execution when running with SCONE, has a binary size of less than 1 \textit{MB}.
The default and intended size of cache available for Memcached is 64 MB, which is lower than a typical EPC size of 90 MB.
It is therefore possible that Memcached does not exceed the EPC during the benchmarks. 
The size of the classification binary with SCONE, on the other hand, was 1.2 MB, a required library needed 15 MB, and the model that is required for classification took another 163 MB. 
Finally, the image needed 1 MB of space.
A check with the Linux \texttt{time} tool showed that the classification process required about 330 MB in main memory for classification, regardless of the number of images.
This led to more paging because it exceeded the typical EPC size.
The results of a microbenchmark conducted in \cite{arnautov2016scone} can be seen in Figure \ref{fig:scone-microbenchmark}.
When the allocated memory is larger than the EPC size, the performance of reads and writes severely degrades.
The same effect could have led to reduced performance when running TensorFlow Lite.

\myparagraph{\#3: Effect of heap size}
The amount of configured \texttt{SCONE\_HEAP} did not significantly affect performance, as long as it is set to the required minimum of about 330 MB or higher. 
This variable controls how much heap memory SCONE allows the application to allocate.
In general, applications may gain performance with higher available main memory, when the application can scale with main memory.
Examples of scaling with memory include applications that use more aggressive caching or launch additional threads.

This common law is invalid in the SGX environment, because of limited EPC size.
A filled EPC requires swapping to unprotected main memory, when more pages are to be allocated, which again requires costly calculations in order to protect the data.
Applications that scale with the amount of available memory may benefit from having virtually less available memory, because using more main memory potentially leads to more EPC swapping operations.
When the benefits of having less EPC page swaps outweigh the benefits of having more available main memory, the user should configure the heap size available for the SCONE application to be artificially smaller than the actual available amount.

\myparagraph{\#4: Effect of file system shield}
Using the file system shield had minimal influence on the performance of the classification process, as can be seen in Table \ref{tab:classification_fss}.
In simulation mode, \projectname with file system shield took about 1\% longer for training, whereas in hardware mode, the difference was 2\%.

The shield uses Intel-CPU-specific hardware instructions for performing cryptographic operations.
These instructions can reach a throughput of up to 4 GB/s, while the model is about 150 MB in size.
This leads to a negligible overhead on the startup of the application only. 
\begin{table}[]
    \centering
    \commentfontsize
    \begin{tabular}{|l||l||l|}
    \hline
    & simulation mode & hardware mode \\ \hline
    \hline
     time (s) w/o fss & 2,157 & 5,875 \\ \hline
     time (s) w/ fss & 2,163 & 5,990 \\ \hline
    \end{tabular}
    \caption[\projectname classification with file system shield]{Comparison of classification times of \projectname for 1,000 images with SCONE in simulation and hardware mode, and with and without file system shield. The heap size was kept at a fixed value.}
    \label{tab:classification_fss}
\end{table}
%
\subsection{Performance of the Training Phase} \label{sec:training-evaluation}
Next, we evaluate the performance of \projectname when training a model.
We answer the question of how costly additional threads are, both at application and SCONE level, and the effect of the maximum heap size granted for the application.
We made separate measurements with the file system shield enabled and disabled to isolate the performance impact of it.
\subsubsection{Data set} \label{sec:training-data-set}
The model we trained was the convolutional neural network we also used for demonstrating that training with \projectname works in general.
The data set was again Cifar-10.
Both items are described in section \ref{sec:tensorflow-full-usage}.
We trained the model up to 10,000 steps.
Each step consisted of forward passing 128 images and adjusting the weights for each image through backward passes. 
Each epoch consisted of 50,000 images in total, which means we trained the model over 25 times with each image available for training.
Since the images were perturbed, training on the same image multiple times still improved accuracy. 

\subsubsection{Results} \label{sec:training-results}
The results of the measurement when training 10,000 step in total can be seen in Figure \ref{fig:training_res}.
The precision we reach after training was about 80\%.
Training longer, the model can reach up to 87\%.
In this section, we describe different aspects and implications of the results.
\iffalse 
\begin{figure*}
    \centering
    \begin{subfigure}[b]{0.4\textwidth}
        \includegraphics[width=\textwidth]{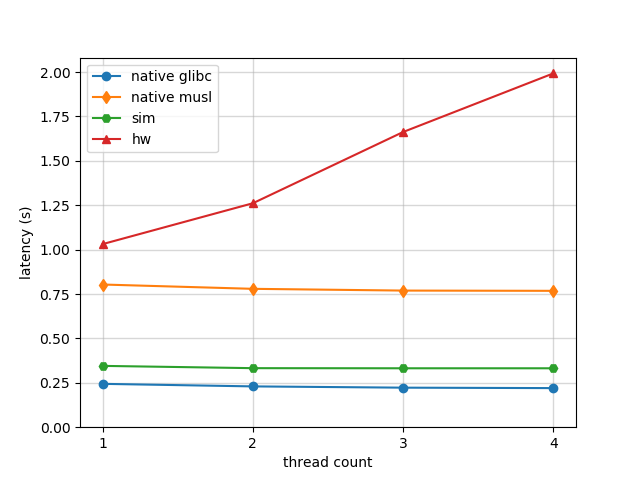}
        \caption{Latency per step when the available heap size is fixed to 700 MB.}
        \label{fig:training_700m}
    \end{subfigure}
    \begin{subfigure}[b]{0.4\textwidth}
        \includegraphics[width=\textwidth]{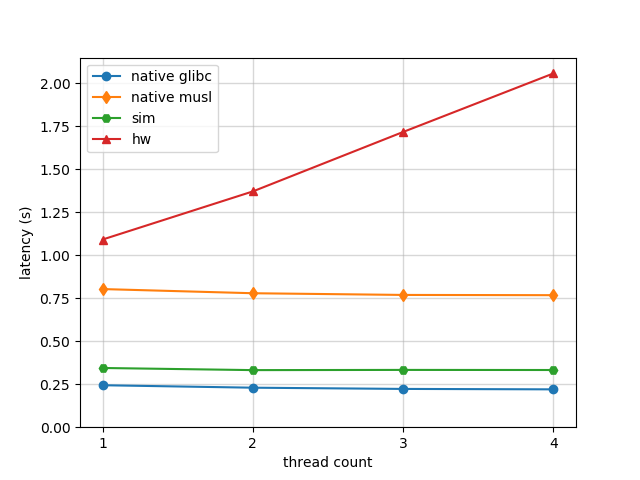}
        \caption{Latency per step when the available heap size is fixed to 2 GB.}
        \label{fig:training_2g}
    \end{subfigure}
    \begin{subfigure}[b]{0.4\textwidth}
        \includegraphics[width=\textwidth]{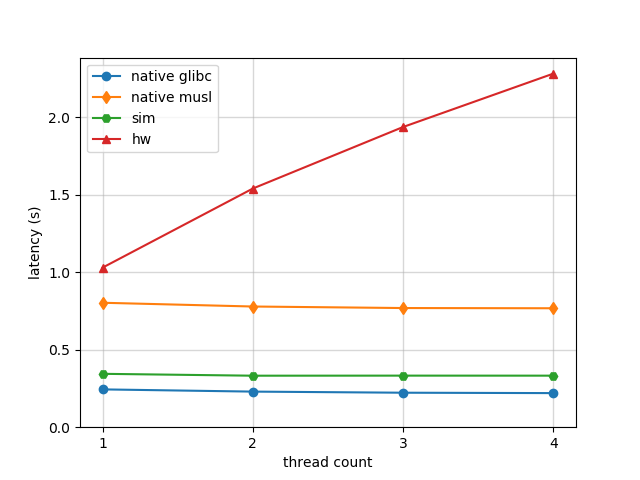}
        \caption{Latency per step when the available heap size is fixed to 8 GB.}
        \label{fig:training_8g}
    \end{subfigure}
    \caption[\projectname training latency]{Latency of the training process for different available heap sizes for TensorFlow. 
            In a single step, 128 images are passed forward in the model, the loss and the gradients are calculated, and the weights and biases are adjusted accordingly, meaning the model is trained to recognize the images.
            We also vary the thread count for the gradient calculation.}
    \label{fig:training_res}   
\end{figure*}

\else

\begin{figure*}
    \centering
        \includegraphics[width=0.8\textwidth]{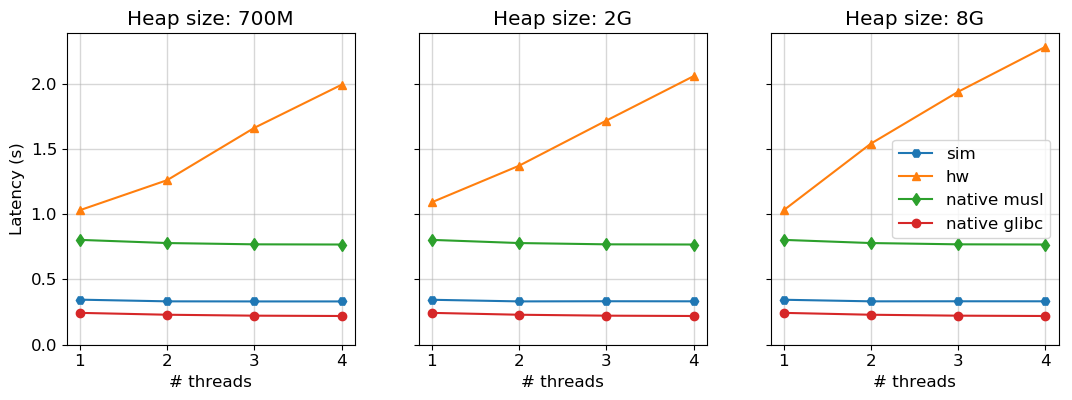}
    \caption[\projectname training latency]{
        Latency of the training process for different available heap sizes for TensorFlow. 
            In a single step, 128 images were passed forward in the model, loss and gradients are calculated, and the weights and biases are adjusted accordingly, meaning the model was trained to recognize the images.
            We also varied the thread count for the gradient calculation.}
    \label{fig:training_res}
\end{figure*}
\fi

\myparagraph{\#1: Latency}
The results of the latency measurements have similarities to the results observed during the classification process in section \ref{sec:tflite-results}.
Deviation of the measurements is low.
Running with SCONE in hardware mode roughly caused a fourfold decrease in performance.
This is an additional factor compared to the training step.
The difference between native musl and SCONE in simulation mode was also much higher than during classification.
The additional performance overhead implied by SCONE may stem from the overall worsened performance of musl when training.
When set to SGX simulation, system calls are handled in userspace, what makes the performance of this approach almost on par with native glibc.
%

\myparagraph{\#2 Effect of heap size}
Setting SCONE heap size to values higher than what is minimally required caused only small latency increases in total. 
\projectname did not scale to available memory.
The minimally required heap size is therefore also the maximum heap size used at any given time. 
We confirmed this by comparing the minimally required heap size that is required for training with SCONE, with the memory allocated by the process in total when no restrictions on heap size were made.

Additional heap size is still allocated and reserved by SCONE.
While this does not directly affect the total count of pages required in the EPC at the same time, it may still lead to decreased performance because of memory fragmentation.
Memory fragmentation causes increased page swapping, when pages are not used to full extent, but instead only partially filled.
When more pages are available, the allocator is more likely to choose pages in a fashion that causes fragmentation.
We conclude that when using \projectname, we should always aim to find the minimally required heap size, which depends on the individual model served.
%

\myparagraph{\#3: Effect of thread count}
The number of threads we varied is the number of threads dedicated to forward passing tensors and updating weights.  
The application still required two additional threads to handle reading images from disk, and for handling queues.

We can tell from the marginal performance improvement in Figure \ref{fig:training_res} that the pure training task is scantly parallelizable on a single machine.
Available resources were already effectively used.
The performance gain from multiple threads when training was consistent, but negligible.

We did learn however that multithreading is possible in general with \projectname, and from the sharp increase in latency we could also estimate the costs for additional threads in \projectname.

\myparagraph{\#4: Effect of file system shield}
\begin{table}[]
\commentfontsize
    \centering
    \begin{tabular}{|l||l|l|l|l||l|l|l|l|}
    \hline
     & \multicolumn{4}{c||}{simulation mode} & \multicolumn{4}{c|}{hardware mode} \\ \hline
    \hline
    thread \# & 1       & 2       & 3       & 4       & 1        & 2        & 3        & 4  \\ \hline
    \hline
     time (s) w/o fss & 3,45 & 3,32 & 3,32 & 3,32 & 10,31 & 12,61 & 16,62 & 19,93 \\ \hline
     time (s) w/ fss & 3,50 & 3,35 & 3,36 & 3,35 & 10,29 & 12,73 & 16,87 & 20,05 \\ \hline
    \end{tabular}
    \caption[\projectname training with file system shield]{Comparison of Cifar-10 training times of \projectname with SCONE in simulation and hardware mode, with different thread counts, and with and without file system shield. The heap size was kept at a fixed value.}
    \label{tab:training_fss}
\end{table}
The measurement results when running with the file system shield enabled are almost identical to the results when not using the file system shield.
We compare both values in Table \ref{tab:training_fss}.
The results when running with file system shield are on average insignificantly higher than when running without file system shield.

Multiple factors hide potential overheads of the file system shield.
As explained in section \ref{sec:tflite-results}, SCONE can reach a throughput of up to 4 GB/s for cryptographic operations, whereas model and data were 150 MB in size, respectively. 
Furthermore, the data was concurrently prepared for the neural network, meaning potential latency may have been hidden in concurrency, because all decryption was performed by the thread filling the queue.
The data queue was usually filled completely, meaning the training calculations dominated reading and decrypting images.

\section{Related Work}
\label{sec:related}

In this section, we summarize the related work about secure machine learning, and shielded execution based on Intel SGX. 

%
%
%
%
Early work on preserving privacy for data mining techniques have relied on randomizing user data \cite{agrawal2000privacy,du2003using, PrivApprox2017}.
These approaches trade acurracy for privacy.
The work of Du et al. \cite{du2003using} includes a parameter that allows making a trade-off between privacy and accuracy.
The used algorithms aim to provide privacy preserving collection of data, and do not protect the results themselves in the cloud, nor do they secure the classification phase. Further, we target to provide the same accuracy level as the native execution.

An approach reaching the same accuracy as unprotected variants is to perform machine learning on encrypted data.
Bost et al. \cite{bost2015machine} developed protocols to perform privacy preserving classification.
While this can protect the privacy of the users of a classification service, it does not cover training as in \projectname.

Graepel et al. \cite{graepel2012ml} developed machine learning algorithms to perform both training and classification on encrypted data.
The solution is based on the properties of homomorphic encryption.
A homomorphic encryption scheme allows operations, such as multiplication and addition, on encrypted data, so that the result can be decrypted by the owner of the private key to yield the same as when performing the operation on the plaintext data. However,   homomorphic encryption schemes provide restrictive compute operations, and incur high performance overheads. 

Shielded execution  provides strong security guarantees for legacy applications running on untrusted platforms~\cite{Baumann2014}. Prominent examples include Haven~\cite{Baumann2014}, SCONE~\cite{arnautov2016scone}, Graphene-SGX~\cite{tsai2017graphene}, Panoply~\cite{shinde2017panoply}, and Eleos~\cite{Orenbach2017}.  Our work builds on the SCONE framework.
 
 Recently, there has been a significant interest in designing secure data analytics systems based on shielded execution. For instance, VC3~\cite{Schuster2015} applies SGX to the domain of big data processing by applying it to the Hadoop MapReduce framework. Along the same lines, Moat~\cite{sinha2015moat} proves confidentiality of enclave programs.  To this end, Moat applies theorem proving and information flow analysis. Opaque~\cite{Opaque} uses Intel SGX to provide oblivious computing to a secure distributed data analytics applications. Likewise, Ryoan~\cite{ hunt2016ryoan} provides a distributed sandbox for untrusted computation on secret data leveraging Intel SGX. EnclaveDB~\cite{enclavedb} is a shielded in-memory SQL database. SGXBOUNDS~\cite{kuvaiskii2017sgxbounds} provides a lightweight memory safety techniques for SGX-based enclaves. 
 In the domain of network data processing, \textsc{Slick}~\cite{slick} and ShieldBox~\cite{shieldbox} use SGX to build a secure middlebox framework for high-performance network processing.   
 In the domain of storage, Pesos~\cite{pesos} focuses on secure data storage using a combination Intel SGX and Kinetic storage. Speicher~\cite{speicher} presents a secure LSM-based KV store using shielded execution. Among all of the recent work, the work from Ohrimenko et al.~\cite{ohrimenko} is the most relevant for \projectname, where they leveraged Intel SGX to secure specific machine learning operators. In contrast to work  from Ohrimenko et al.~\cite{ohrimenko}, we present the first generic machine learning framework based on the widely-used TensorFlow framework, which can support a wide-range of unmodified existing TensorFlow applications. 

Currently, \projectname  does not make use of GPUs to deploy TensorFlow operators since they do not provide a TEE. In this space, Graviton~\cite{graviton} recently proposed hardware extensions to provide a secure  environment on GPUs. We plan to leverage Graviton's extensions for deploying \projectname applications on GPUs. 

Typically, machine learning applications, by their nature, are error-tolerant~\cite{approxann}. Meanwhile, approximate computing has recently emerged as a design paradigm that allows us to  make a trade-off between the out quality, performance, and computing resources for data analytics~\cite{incapprox-www-2016, streamapprox-middleware2017, streamapprox-tech-report, PrivApprox2017, privapprox-tech-report,  approxiot-icdcs-2018, approxiot-tech-report, approxjoin-socc-2018, approxjoin-tech-report, approx-thesis}. Therefore, a promising approach to further improve the performance of \projectname, i.e., to reduce the computation overhead inside enclaves, is to apply approximate computing techniques.

\section{Conclusion} 
\label{sec:conclusion}
In this paper, we introduced \projectname,  a secure TensorFlow-based machine learning framework leveraging the hardware-assisted trusted execution environment (TEE) based on Intel SGX.  More specifically, we have presented the design of \projectname based on the integration of TensorFlow with the SCONE shielded execution framework. We have implemented \projectname as a fully functional system supporting many useful optimizations to overcome the architectural limitations of Intel SGX in the context of building a secure machine learning system. \projectname  supports both training and classification phases while providing all three important design properties for the secure machine learning workflow: {\em transparency},  {\em accuracy}, and {\em performance}. Our evaluation shows that \projectname incurs reasonable performance overheads, while providing strong security properties against a powerful adversary. 
\balance
\bibliographystyle{abbrv}
\bibliography{main}

\end{document}